# The ABC of digital health:
# A framework for translating digital health interventions into real-world applications

David Grüning*[1,2,3]
Vincent Beermann[4]
Jan Enkmann[4]
Rose Hoch[1]
Ralph Hertwig[2]
Paul Schmiedmayer[1]

1: Stanford University, Palo Alto, US
2: Center for Adaptive Rationality, Max-Planck Institute for Human Development, Berlin, Germany
3: University of Cambridge, Cambridge, UK
4: Hasso Plattner Institute, University of Potsdam, Potsdam, Germany

*Corresponding author: gruening@stanford.edu, dgruening@mpib-berlin.mpg.de, david.gruning@mrc-cbu.cam.ac.uk

---

## Acknowledgements

DG's work is funded by Huo Family Foundation and Stanford's Center for Digital Health. DG has ongoing research projects in collaboration with the one sec app. PS's work is funded by Stanford's Center for Digital Health. RH has been funded by the Deutsche Forschungsgemeinschaft (DFG, German Research Foundation) – HE 2768/11-1.

## Contribution statement

Framework conceptualization: DG, VB, JE. Theoretical formalization: DG. Literature synthesis: DG, VB, JE, RoH. Writing original draft: DG. Visualizations: DG. Revised the article: DG, VB, JE, RoH, RH, PS. Approved the submitted version for publication: DG, VB, JE, RoH, RH, PS.

## Abstract

Research-based digital health interventions are often presented as potential solutions for extending health care in the real world. Yet the vast majority of these interventions fails to move beyond controlled studies. Existing frameworks offer valuable guidance for intervention development and testing, but provide less concrete support for translating these evidenced intervention mechanisms into sustained real-world applications. This paper introduces the ABC framework, referring to Accessibility, Buildability, and Continuity, as a practical model for a successful translation. Accessibility captures whether diverse users can find, understand, and begin using an application with minimal friction. Buildability refers to the development of an app that supports the iteration, integration, and personalization of features. Continuity describes both sustained user engagement and the operational capacity to maintain an application over time without disproportionate increases in cost, infrastructure, or human support. Different combinations of the ABC-dimensions make an application scalable (AB), automated (BC), and adherent (AC). By linking design decisions to these features, ABC offers a shared language for researchers, designers, and policymakers seeking to build or evaluate digital health interventions that work beyond trials and are viable applications in everyday life.

## Introduction

Digital health interventions have grown from being considered interesting experiments to essential parts of people's everyday life. Ranging from mental wellbeing to chronic disease management, health applications are increasingly suggested as scalable tools to extend care, reduce health disparities, and support long-term behavior change (Nahum-Shani et al., 2018). Their promise lies not only in their reach but also in their potential to operate independently of constant human input (Mohr et al., 2017). Yet while enthusiasm for these digital interventions is widespread in academic research, actual implementation in real life often falls short of expectations. Applications are predominantly built for intervention testing but rarely used beyond answering a research question. Features are developed but not implemented in actual digital environments for prolonged use. We argue that at the heart of this failure lies a design problem: the lack of practical frameworks that guide the translation of evidenced digital health interventions into real applications.

## Existing models

The gap between academic promise and real-world performance has prompted growing interest in frameworks that explicitly guide not just the evaluation but the design of actionable digital health interventions. As shown in Table 1, multiple conceptual models exist to guide digital intervention design, ranging from implementation science approaches to behavior change taxonomies. However, these models often lack the technical specificity required to inform software development or automation logic. For instance, while the DEDHI framework supports end-to-end intervention planning, it does not fully address operational buildability or modular reusability required for scale (Kowatsch et al., 2019). Similarly, while behavior change frameworks like the Behavior Change Wheel offer strong psychological foundations, their application to app architecture and feature logic is limited (Michie et al., 2011). Frameworks such as PICOTS-ComTeC can help standardize how digital health interventions are reported and evaluated, especially for outcomes research (Zrubka et al., 2024), while, yet other perspectives (Jahnel et al. 2025; Wienert et al., 2022) emphasize the public health lens in digital design, considering structural, contextual, and legal variables throughout intervention development. All of these models still struggle to connect high-level policy or health-system concerns with on-the-ground development constraints such as cost of feature iteration, backend integration, or sustaining user engagement over time.

Existing work tends to treat scalability as a downstream outcome rather than a design constraint. Many interventions developed in research contexts prioritize experimental control over real-world usability, leading to solutions that are well-intentioned but poorly scalable in the digital wild (Mohr et al., 2017; Yardley et al., 2015). This observation is echoed in work by Zamboni et al. (2019), who argue that many trials still treat scalability as something to assess after efficacy is established-despite calls to design for scale from the outset. Similarly, Milat et al. (2020) present the Intervention Scalability Assessment Tool (ISAT) as a practical way to evaluate scale readiness, but its emphasis lies more in assessment than in informing feature development or architecture choices.

Researchers are increasingly modelling automation in terms of user-facing continuity and backend independence. For example, the APLES framework proposes a system for automated planning of intervention content, using rules-based logic to generate engagement structures dynamically (James et al., 2024). This moves the automation conversation from abstract infrastructure into implementable tooling. Complementing this, O'Raghallaigh and Adam (2017) present a theory-based framework for designing digital health interventions, called Digitally

Based Change Interventions (DBCI), that draws from cognitive science and implementation theory to inform the design of just-in-time interventions in digital applications. While not automation-specific, their framework provides needed granularity in translating theoretical constructs into design decisions that personalize themselves to the user.

Grüning et al. (2024) introduced a conceptual pyramid to characterize effective digital health interventions, identifying four key attributes: digital, scalable, automated, and tested. To the best of our knowledge, their model is the first one that explicitly elevates scalability and automation to central design goals rather than observed outcomes of a digital intervention. However, the model remains diagnostic rather than generative as it outlines what a digital intervention looks like that scales and is automated, but it falls short of principles to engineer this scalability and automation to create a real-world application.

In the present paper, we introduce the ABC framework to address this gap. The framework provides a bridge from high-level principles of intervention research (Grüning et al., 2024) to on-the-ground application decisions as they interest practitioners and users (Grüning & Kamin, 2026; Menking et al., 2025).

**Table 1**

*Existing frameworks for digital health intervention design, relative to the ABC dimensions.*

| Framework | Source | Core focus | Access. | Build. | Contin. | Key gap relative to ABC |
|---|---|---|---|---|---|---|
| **DEDHI** | Kowatsch et al., 2019 | End-to-end intervention planning & evaluation | ◐ | ◐ | ◐ | Lacks operational specificity for buildability or modular reusability at scale |
| **Behavior Change Wheel** | Michie et al., 2011 | Psychological / behavioral mechanism design | ◐ | ○ | ○ | Strong on mechanism, limited translation to app architecture or feature logic |
| **PICOTS-ComTeC** | Zrubka et al., 2024 | Standardized reporting & outcome evaluation | ○ | ○ | ◐ | Built for reporting/ evaluation, not feature-level design decisions |
| **Public health lens** | Jahnel et al., 2025; Wienert et al., 2022 | Structural, contextual, and legal dimensions of digital health | ◐ | ○ | ○ | System-level context, not on-the-ground development constraints |
| **ISAT** | Milat et al., 2020 | Scale-readiness assessment | ○ | ◐ | ○ | Assesses scalability after the fact rather than guiding architecture choices |
| **APLES** | James et al., 2024 | Rules-based automated content planning | ○ | ◐ | ◐ | Automation-specific; doesn't address accessibility or broader deployment |
| **DBCI** | O'Raghallaigh & Adam, 2017 | Theory-based design of just-in-time interventions | ◐ | ◐ | ◐ | Personalization granularity, but no integrated model spanning all three concerns |
| **Four-attribute pyramid** | Grüning et al., 2024 | Diagnostic model: digital, scalable, automated, tested | — | — | — | Diagnostic rather than generative, names the target state, not how to build toward it |

*Note.* ● = central focus, ◐ = partially addressed, ○ = not addressed, — = not organized around comparable dimensions. Access. = Accessibility, Build. = Buildability, Contin. = Continuity.

## The ABC framework

ABC refers to Accessibility, Buildability, and Continuity. In contrast to purely conceptual taxonomies, this framework focuses on operational features: inclusive design (A), modular and adaptable development (B), and mechanisms for long-term engagement (C). Each dimension addresses a distinct layer of intervention design, i.e., accessibility focuses on user experience, buildability on the developer infrastructure, and continuity on retention and feedback mechanisms. As such, the ABC dimensions are not abstract ideals but we understand them as design levers that can be implemented, their effects measured, and their implementation iteratively developed within a real-world application (Pagliari, 2007). In the following, each of the three dimensions is outlined alongside strategies for its implementation and evaluation as well as helpful design principles. For the design principles, we draw on commercial apps as existence proofs of translatable design patterns, not as evidence-based interventions themselves. Table 2 summarizes these characteristics for each ABC dimension.

We hasten to note three clarifications here. First, the ABC framework is not a framework for designing intervention mechanisms themselves. That is, it is not intended to replace behavior change or evidence-generation frameworks such as the ones mentioned above. Rather, it complements them by focusing on deployability, scalability, and sustained real-world use of identified mechanisms. ABC assumes that the underlying intervention mechanism is evidence-informed. It then guides someone on successfully translating the evidence-based mechanism into a sustainable application. Second, smartphone apps are the focus of the present framework for a specific reason: they represent the most prominent and widely distributed digital platform for health interventions available today. Unlike interventions embedded in clinical systems, hardware, or institutional environments, apps can be deployed directly to users without institutional gatekeeping. But this also raises the stakes, as poor but intuitive design spreads just as easily as good design. Further, we argue that app-based interventions must not only work, they must also be used, kept, and relied on by real people. In Table A1 (Appendix), we list all smartphone applications referenced as examples in the following chapters of this paper. Third, to our knowledge, few if any evidence-based interventions have achieved the scale and sustained use of the commercial applications discussed below, that are not clinical applications, a scarcity that itself illustrates the translation gap motivating this paper. We therefore draw design principles primarily from prominent commercial applications, several of which build on well-established behavior-change mechanisms even without formal efficacy trials, and use one sec, validated in natural, controlled trials (Grüning et al., 2023; Haliburton et al., 2024; Hansen et al., 2026) and now in real-world use, as the closest available case of a tested intervention that scaled.

**Table 2**

*Dimensions of the ABC framework.*

| Component | Description | Implementation | Evaluation | Design Principle Examples |
|---|---|---|---|---|
| **Accessible:** Inclusive design and minimal friction | The ease with which diverse users can find, understand, and begin using an intervention. Encompasses intuitive interfaces, frictionless onboarding, and technical flexibility across devices and contexts. | Following human-centered, participatory design methods including deep user research to surface hidden needs, and generalize to underserved user groups. | Combine quantitative metrics (e.g., onboarding completion and early dropout rates, adoption among low-resource users) with qualitative indicators (e.g., user-reported sense of agency and control). | ● *Accessible entry:* Offer meaningful free tiers that deliver core value immediately without payment, allowing users to experience behavior change before committing financially.<br>● *Rapid value delivery:* Enable meaningful engagement within 5 minutes of download through streamlined onboarding, minimizing cognitive load when establishing new behaviors.<br>● *Progressive disclosure:* Hide complexity behind simple interfaces, revealing features as users mature in engagement, scaffolding rather than overwhelming. |
| **Buildable:** Modularity and low-friction development | The ease with which an intervention can be developed, iterated, and scaled. Encompasses modular architecture, rapid prototyping, API ecosystems, and transparent user customization options. | Developer side: Buildable through low- or no-code environments, agile development, and iterative design in cross-functional teams.<br>User side: adjustable notification or interface settings. | Developer side: Average feature development duration, frequency of updates, integration with other platforms.<br>User side: Engagement with adjustable features (quantitative) and subjective evaluation of unambiguity and usefulness (qualitative). | ● *Modular architecture:* Structure content and features as independent modules that can be mixed, matched, and updated without system-wide changes, enabling both personalization and rapid iteration.<br>● *User-generated scaling:* Leverage users to create value for other users, building network effects while reducing content creation costs.<br>● *Ecosystem integration:* Build API-first integration capabilities with complementary services, fitting into existing technology ecosystems rather than creating walled gardens. |
| **Continuous:** Long-term engagement and feedback | The ability to maintain user engagement over time in sustainable, non-intrusive ways. Encompasses psychological reinforcement, low-maintenance integration into daily routines, and passive value delivery. | (Self-)Nudging strategies (e.g., defaults), JITAI features, or habit loop design to facilitate usage over longer time frames. | Retention rates (i.e., 30, 60, or 90 day user rates), direct outcome behavior assessments (e.g., reductions in screen time), or self-reports on habit strength or target behavior maintenance. | ● *Strategic friction and progress:* Introduce deliberate pauses for unwanted behaviors and visible, loseable progress for desired behaviors, leveraging loss aversion while avoiding punitive mechanics that shame users.<br>● *Ambient presence:* Deliver value even when not actively used through background monitoring and automatic interventions, recognizing that continuity does not require constant active engagement.<br>● *Optional social accountability:* Add opt-in community features that transform individual behavior change into social commitment, leveraging peer accountability while avoiding mandatory exposure that may embarrass or create unhealthy competition. |

**A - Accessible: Inclusive design and minimal friction**

Accessibility in digital health extends far beyond screen reader compatibility or text size adjustments. In this framework, accessibility refers to the ease with which real users can find, understand, and begin using a digital health intervention. A truly accessible application reduces entry barriers to the minimum.

This begins with inclusive user experience design: interfaces that are visually intuitive, cognitively lightweight, and emotionally neutral. Users should not feel overwhelmed, condescended to, or confused. Effective apps offer onboarding that is frictionless, which means avoiding long sign-up processes, excessive permissions, or unexplained medical language that might alienate non-expert users. Accessibility also includes technical flexibility, such as functionality under low-bandwidth conditions or limited data environments, which is crucial for inclusion across socioeconomic and geographic contexts. To create this kind of experience, design teams often rely on human-centered design principles, engaging in deep user research to understand needs and limitations. Participatory design methods (e.g., Grüning et al., 2026) are especially valuable in surfacing hidden frictions and ensuring that the interface empowers rather than frustrates. Usability testing must extend beyond tech-savvy early adopters to underserved groups.

Evaluating accessibility requires a blend of quantitative and qualitative data. Onboarding completion rates and early dropout metrics can reveal initial friction points. Adoption metrics among lower-resource or non-majority users indicate how broadly usable the application truly is. Finally, subjective measures of user agency, such as whether users feel in control of the application, rather than the application controlling them, offer insight into the psychological accessibility of the experience.

In practice, accessibility can be operationalized following at least three design principles. First, successful apps center on economic and cognitive accessibility from first contact, recognizing that users need to experience behavior change before they commit to it financially or psychologically. Hence, they deliver core value immediately without requiring payment or extensive setup. For example, *one sec* offers single-app intervention capabilities without paywalls and even apps that eventually monetize through subscriptions, such as *Headspace* and *Calm*, provide free trials that deliver therapeutic value rather than merely teasing premium features. Hence, accessible apps avoid paywalling core value propositions to create pathways to sustained engagement. Second, leading digital health applications consider temporal accessibility and cognitive load and enable meaningful engagement immediately after download, recognizing that the window for initial value delivery is narrow. *Headspace* delivers a first meditation session within minutes of setup and *one sec* provides instant intervention capability as soon as users select an app to intervene with. Hence, onboarding flows should deliver first value within two to three screens, avoid lengthy tutorials, complex profile creation, or excessive permission requests before users experience core functionality. Third, intervention complexity is introduced progressively. Successful apps simplify complexity with understandable interfaces and reveal their full functionality as users mature in their engagement with the app. For example, *Strava* begins with nudging users toward a single explicit action, recording a run or ride, before gradually introducing new segments like social interaction features, leaderboards, and performance analytics. Similarly, *one sec* starts with a simple breathing pause intervention before revealing additional self-control mechanisms and customization options. This approach reflects that accessibility is not about reducing functionality but about the careful guidance of the users' experience along functionality.

**B - Buildable: Modularity and low-friction development**

Buildability refers to the ease with which an intervention can be developed, iterated, and scaled, both by its creators and by the systems around it. From a developer's perspective, a buildable application has a modular structure, enabling quick prototyping, updates, and integration with other platforms. From a user perspective, a buildable app provides clear, transparent options for customization, allowing it to adapt to individual needs without overwhelming the user.

On the technical side, modularity is central. This means features can be added, removed, or reconfigured without needing to rebuild the entire application. Buildability begins with reusable architectures and shared open-source foundations that reduce duplicated development effort across interventions. Highly buildable applications rely on low-code or no-code environments, reducing dependence on full-stack developers and enabling cross-functional teams to contribute more easily. Open API ecosystems and component-based frontends support this modularity, while also making it easier to integrate with existing systems like wearables, electronic health records, or other applications in the user's digital ecosystem. User-facing buildability means offering flexibility without confusion. Effective apps allow users to select and personalize features that matter to them, such as choosing which notifications to receive, adjusting motivational messages, or changing interface elements like themes or sounds. However, these options must be accompanied by transparency. Users should be able to understand what each feature can and cannot do, and how their feature choices affect their experience. This approach enables teams to respond to user needs in real time and to roll out improvements incrementally.

Evaluating buildability can be done through several indicators. These include the average time to prototype or release a new feature, the frequency of updates over time, and the degree of integration with external platforms. On the user side, buildability is reflected in whether users understand what features are available to them, how much users customize their features, and how often users engage with these features.

We argue that successful real-world applications often implement some of three architectural patterns that enable both rapid iteration and user personalization at scale. First, successful apps structure content and features as independent modules that can be mixed, matched, and updated without requiring system-wide changes. *Calm* creates individual sleep stories as discrete units that can be added to the library without modifying the playback system. This modularity serves dual purposes: it enables personalization for users who can select and combine relevant features, and it facilitates rapid iteration for developers who can update individual components without rebuilding entire systems. The alternative, i.e., apps built with monolithic structures, can move quickly in early development but slow dramatically as feature complexity grows, ultimately limiting their ability to adapt to user feedback and their changing health needs. Second, apps can leverage users to create value for other users which reduces content creation costs while building network effects. *Strava's* segment system allows any user to define a route segment that becomes a challenge for the entire community, creating millions of user-generated competitions. *MyFitnessPal* built a database of over 14 million foods largely through user contributions, a resource that would have cost millions to develop internally. This principle addresses a fundamental scaling challenge: internally generated content cannot grow as fast as user demand. However, in a user-generated model, each new user not only potentially adds value for all existing users but these network effects are also self-reinforcing, more *Strava* users create more segments, making the app more valuable for everyone, which attracts more users. However, user-generated scaling requires careful design to ensure quality and prevent abuse, e.g., by including moderation mechanisms, reputation systems, and a curation that maintains value while distributing creation effort. Third, successful apps focus on integration rather than isolation. Apps that achieve lasting impact build integration capabilities with complementary services, multiplying their value without proportional development effort. *Strava* integrates with over 30 fitness devices and platforms, automatically importing activity data from Garmin watches, Peloton bikes, and dozens of other sources. This API-first approach fits behavior change tools within users' existing technology ecosystems rather than requiring them to abandon familiar tools or manually transfer data. The integration creates unexpected value, *Strava's* open API has resulted in an ecosystem of third-party applications for training analysis, route planning, and social features that *Strava* itself did not build.

**C - Continuous: Long-term engagement through customization and feedback**

Continuity refers to an app's ability to maintain user engagement over time in a sustainable, non-intrusive way. While accessibility gets users in the door and buildability keeps the system agile, continuity ensures that behavior change actually lasts. This dimension is not just about re-engagement strategies, it is about designing interventions that embed themselves meaningfully into users' daily lives.

This starts with psychological reinforcement. For instance, applications can offer some form of visible progress feedback, whether it is streaks, milestones, or personal reflections (e.g., Hillegass et al., 2025), to signal that continued use is worthwhile. Self-tracking features become powerful when they go beyond raw data to help users reflect on their own patterns, like habits or triggers. Crucially, continuity also relies on low-maintenance integration: once the application is set up by the user, it should not require excessive interaction. Instead, it should offer value passively or with minimal effort. Effective continuity strategies often draw on behavioral economics, such as nudging or default settings that support beneficial behaviors (Johnson & Goldstein, 2003; Thaler & Sunstein, 2008). Just-in-time adaptive interventions (adjust timing and content based on user context; Klasnja et al., 2015; Nahum-Shani et al., 2018) and habit loops (embed the app into predictable behavioral cycles; Lally et al., 2010; Wood & Neal, 2007) could further increase long-term use.

Evaluating continuity involves tracking retention over time besides the number of downloads or early users. Metrics such as 30-, 60-, and 90-day active user rates offer a clearer picture of whether an app sustains value and where it might fall off over time. Other indicators include self-reported habit strength, maintenance of target behaviors (e.g., reduced screen time, improved step counts), and the frequency of meaningful feedback events (e.g., user-initiated reflection or journaling).

Several design principles can be derived from successful apps. First, they provide direct feedback about the intervention's effectiveness and the user's progress, allowing users to experience the results from their interactions with the app. For example, on its front page *one sec* provides statistics about the number of abandoned interactions with unwanted apps and translates it into the overall time the user has individually saved thanks to the interventions provided. *Headspace's* "run-streak" feature adds up users' meditation practice which resets to zero when skipping a daily practice, making abandonment feel like loss. Importantly however, successful implementations of loss aversion avoid punitive mechanics that shame users. *one sec* permits users to dismiss interventions to maintain autonomy. The balance is delicate: losable streaks are highly effective in fostering motivation and sustaining engagement while users feel they can attain them. They can backfire once a streak is broken, a limitation that can, however, be attenuated by the option to restore the lost streak (Silverman & Barasch, 2023). One caveat cuts across these reinforcement mechanics. Extrinsic motivators such as streaks, points, and competition reliably raise short-run engagement, but a large body of experiments shows they can crowd out the intrinsic motivation that sustains a behavior once the reward (or the app) is removed (Deci et al., 1999); gamified social-comparison features in particular, such as leaderboards and competitive streaks, have been found to lower intrinsic motivation and satisfaction over time rather than raise them (Hanus & Fox, 2015). For a framework whose aim is durable, post-trial use, this matters: continuity built purely on extrinsic hooks risks producing engagement with the app rather than maintenance of the target behavior. We therefore treat ambient value and self-reflective feedback, which return insight and control to the user, as the more durable continuity levers, and recommend that extrinsic mechanics be designed to fade as the underlying behavior becomes self-sustaining. A second principle relates to the ambient presence of apps, the ability to deliver value even when not actively used. This challenges a common misconception that continuity requires constant engagement. *Calm's* sleep timer provides value during unconscious hours, requiring no active participation once set. *one sec's* interventions trigger automatically based on user behavior patterns, shaping choices without requiring users to remember to activate protections. Users may not open *one sec* for weeks but continue receiving value from its interventions once pre-set. From a design perspective, ambient presence requires thoughtful notification strategies, background processing capabilities, and clear value delivery even during periods of low interaction. It also requires transparent communication about what the app does when not actively used, addressing privacy and autonomy concerns (Kidman et al., 2024). The third principle recognizes the power of social accountability. Community features transform individual behavior change into social commitments that outlast personal motivation. *Strava's* kudos system provides social reinforcement for completed activities, creating lightweight but meaningful acknowledgment from peers. *Habitica's* party system introduces peer accountability where one person's failure to complete tasks affects the group's progress in the game. The impact of such social features can be significant: *Strava* reported a 59% increase in run clubs in 2024, with users completing 40% longer activities in group settings compared to individual efforts (Strava, 2024). However, social features must be carefully implemented and the most effective implementations allow users to choose their level of social engagement, from completely private use to full community participation. *Strava* allows users to mark activities as private and *Habitica* supports both solo and party play. This flexibility recognizes that social accountability might be powerful for some users in some contexts but counterproductive for others (Zhu et al., 2021). The design implication is that social features should be additive rather than core, enhancing continuity for those who want them without creating barriers for those who do not.

We hasten to note that continuity extends beyond sustained user engagement. Operational continuity is equally important and refers to an application's ability to remain viable as usage grows over time. This includes sustainable cost structures, scalable technical infrastructure, maintainable software architectures, and the avoidance of support models that require human resources, which increase proportionally with the number of users. Digital health interventions cannot just fail because users disengage, but because the economic or organizational demands of maintaining the intervention become unsustainable with an increasing amount of users. Accordingly, continuity should be understood not only as the persistence of user engagement, but also as the long-term operational capacity of an intervention to deliver value without requiring unreasonably increasing resources.

## ABC as an operationalisation of scalable and automated interventions

As already noted in the above introduction of existing models, we argue that scalability and automation are the most desirable qualities to translate digital health interventions into actual applications. But they are also one of the most misunderstood. Often framed as end-stage outcomes of a successful intervention, they are rarely treated as design principles that must be built in an application from the start. The ABC framework operationalizes the latter approach by reframing scalability and automation as the natural results of thoughtful design across the present three key domains: accessibility, buildability, and continuity. In the following, we explain how the ABC components interact to produce these two qualities of an application and a third one: adherence.

### What scalability actually means

Scalability in digital health is not simply about reaching large user numbers. Rather, it refers to an intervention's ability to expand its reach and impact without requiring proportional increases in resources. This includes at least three related dimensions. Intervention scalability refers to the ability to reach and support larger and more diverse user populations. Technical scalability describes the ability of the underlying infrastructure to handle increasing demand without substantial performance losses. Economic and organizational scalability concern whether additional users can be supported without proportionally increasing costs, staffing requirements, or maintenance efforts.

The present framework primarily focuses on intervention and organizational scalability. A scalable application can onboard more users, adapt to new contexts, and sustain higher levels of engagement without requiring extensive redevelopment, specialized personnel, or continuously increasing operational resources. In practice, scalability emerges when two conditions are met: First, the application is accessible to a potentially large subset of a target population. Second, the application is buildable, meaning that it can be extended, maintained, and adapted without excessive technical or organizational burden. Without accessibility, an intervention cannot reach scale, regardless of its technical sophistication. Without buildability, an intervention becomes brittle, overextended, or too costly to sustain as it grows. Scalability, we therefore argue, is the product of accessible and buildable design.

**What automation actually requires**

Automation is often misunderstood as a technical feature - something enabled by AI, algorithms, or backend logic. But in the context of behavior change and digital health, automation is about self-sustaining intervention, that is, systems that continue to deliver value and adapt over time without constant human effort. This kind of automation requires two design aspects. First, the application has to have a buildable foundation that supports logic-driven workflows, data triggers, and modular updates. Second, the application needs a continuous user experience that promotes long-term behavior change through passive support, changes in the environmental architecture, and embedded feedback loops. Importantly, automation does not just reduce developer workload - it reduces the need for ongoing user effort. When users do not need to remember, decide, or restart, automation is successful. When the app adapts without prompting, personalization without fatigue becomes possible. This is the intersection of buildability and continuity.

**Sustained adherence as a third outcome**

While scalability and automation are crucial for extending reach and reducing effort, neither guarantees that users will continue engaging with an intervention. The history of digital health is filled with examples of apps that performed well in controlled trials but quickly lost momentum in everyday contexts because of poor adherence (Eysenbach, 2005). This gap between initial uptake and long-term use highlights the need to treat adherence not as a secondary concern, as is, for example, done by Grüning et al. (2024), but as a core outcome of intervention design. Within the ABC framework, adherence emerges at the intersection of accessibility and continuity. Accessibility ensures that a broad and diverse population can begin using an intervention without encountering barriers such as excessive cost, complex setup, or poor usability (Norman, 2013). Continuity guarantees that once users are onboarded, they have ongoing reasons to stay engaged through features that provide sustained value, adaptive feedback, and integration into daily routines (Yardley et al., 2015). Middleton et al. (2013) show that both early engagement and long-term reinforcement mechanisms are essential: adherence depends as much on first impressions as on sustained support.

Recent evidence underlines this dual requirement further. A 2024 study by Mason et al. proposed a multidimensional framework for measuring adherence, distinguishing between adoption, consistency, duration, and dropout, showing that adherence is not a single outcome but a composite of behaviors across time. Similarly, systematic reviews (Karekla et al., 2019; Montalescot et al., 2024) emphasize that adherence is shaped by both design features (e.g., reminders, personalization, ease of navigation) and user-level characteristics (e.g., health status, social support, comorbidities). In long-term trials, interventions that combined technical support with participant-centered engagement strategies showed markedly higher adherence, even over extended periods of 12 months or more (Wilton et al., 2025). Adherence, therefore, is best understood as the alignment of ease of entry with the likelihood of continuous use. An intervention that is easy to adopt but fails to sustain engagement risks becoming a novelty, producing wide but shallow uptake. Conversely, an intervention that offers strong continuity mechanisms but is difficult to access will never attract a sufficiently broad user base to achieve meaningful impact. Without adherence, scalability yields reach without depth, and automation risks delivering efficiency without actual use. That is, we argue, sustained engagement is not an optional benefit, it is a practical necessity for ensuring that scalability leads to meaningful uptake and that automation delivers real-world health benefits rather than theoretical potential.

**Mapping the ABC**

To clarify the relationships, the ABC framework can be visualized as a triangle where each corner supports a broader goal (Figure 1). As explained above, scalability emerges at the intersection of accessibility and buildability (AB), automation emerges at the intersection of buildability and continuity (BC), and adherence emerges at the intersection of accessibility and continuity (AC). This mapping serves two purposes. First, it means to show why no single pillar of the ABC is sufficient for creating a successful real-world application. For example, a highly accessible app that is not buildable will buckle under growth. A highly automated app that is not accessible will reach no one.

Second, the ABC framework serves as more than a conceptual model. It is a practical diagnostic and planning tool that can guide the design, evaluation, and also funding of digital health interventions. For developers and designers, the framework offers a structured way to interrogate design decisions before a single feature is built. Questions such as “Who can use this easily and independently?” highlight the need for accessibility from the start. Similarly, asking “Can we modify this in a week or add features in a sprint?” directs attention to buildability, ensuring that the product can evolve without excessive technical debt. Reflecting on “Will people still use this six weeks from now, and why?” emphasizes continuity, forcing teams to embed mechanisms that support sustained engagement rather than novelty-driven spikes in use. For researchers, the ABC framework extends the evaluation of digital health interventions beyond short-term efficacy. It highlights whether an intervention can continue delivering outcomes outside controlled conditions, adapt to diverse contexts, and integrate meaningfully into health systems. For funders and policymakers, ABC provides a straightforward checklist for regulatory and investment decisions. After empirically tested, interventions must demonstrate that they are accessible enough to achieve wide reach, buildable enough to be maintained or updated with reasonable effort, and continuous enough to ensure that users actually stick with them. Without affirmative answers to all three, the long-term success of a digital health tool remains questionable.

**Figure 1.**
*Visualization of the interaction of accessibility, buildability, and continuity to allow scalability, automation, and adherence.*

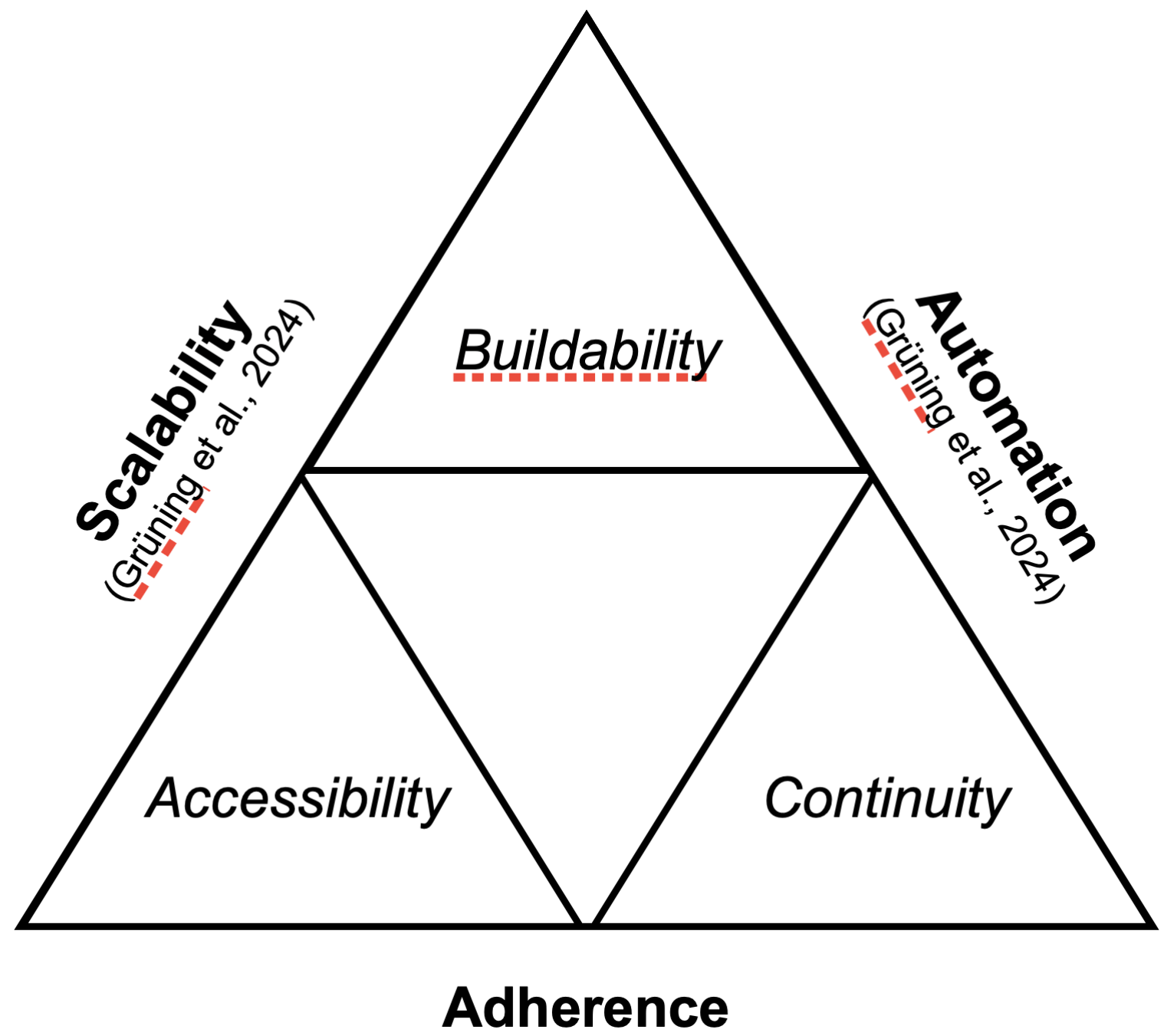

In this way, the ABC framework provides a shared language across technical, behavioral, and strategic domains, hopefully enabling more coherent collaboration between disciplines (Menking et al., 2025). It ensures that digital health tools are not only innovative in theory, but viable in practice.

## Toward experience-driven design

Many digital health interventions begin with a well-documented problem, such as anxiety, sedentary lifestyles, medication non-adherence, or poor sleep. However, the existence of a problem in research does not automatically create demand for an interventive application. Users engage with interventions when they address a need that is experienced in everyday life and perceived as both relevant and changeable (Borghouts et al., 2021; Jardine et al., 2024; Yardley et al., 2015). This distinction is particularly important in academic intervention development. Research projects often begin with theoretically important problems and are optimized for scientific rigor rather than everyday usability. As a result, interventions may address clinically relevant issues while failing to fit naturally into users' routines, priorities, or motivations. A stress-management app, for example, may offer evidence-based features such as breathing exercises or mood tracking, yet still fail if users do not perceive stress as a problem they want to address through an app.

At the same time, it is important to acknowledge that there is little incentive in academic systems to build interventions that are actually used (Menking et al., 2025). Success is measured by publication, grant acquisition, and theoretical contribution, not necessarily or strongly so by application retention, user satisfaction, or public uptake. Until these systems of recognition are shifted, this misalignment will continue, and practitioners will stay rather uninterested in the research on these interventions in the academic spaces. The promise of digital health depends not just on good intentions or solid evidence, but on interventions that users actually want to keep using.

The ABC framework can speak to both issues in the field. First, it encourages a shift from problem-driven to experience-driven design of digital health interventions. Accessibility emphasizes understanding how users perceive and encounter a problem, Buildability enables rapid adaptation to emerging user needs, and Continuity focuses on creating value that remains relevant over time. Together, these dimensions help ensure that interventions are designed not only around documented problems, but around needs that users themselves recognize and want to address. Second, our framework reorients what counts as a “problem worth solving” in digital health. It is not just what shows up in prevalence data or policy reports, it is what shows up in the user’s life and feels possible to change. The success of a digital real-world application not only depends on whether it is grounded in evidence, but rooted in people’s experienced need. This reframing can be achieved through the ABC-framework and helps move the field forward, from building apps about health problems, to building tools that users feel are truly theirs to use.

## Discussion

The ABC framework was developed to refine and ground the often vague goals of scalability, automation, and adherence in digital health interventions to translate them into real-world applications. Where earlier frameworks, such as Grüning et al.’s (2024) four-level pyramid, offered a conceptual direction, ABC makes these three goals actionable. It does so, by translating scalability, automation, and adherence into design-level strategies. Scalability arises not from size alone but from frictionless access (Accessibility) and modular, lightweight systems (Buildability). Automation is not simply backend intelligence, it emerges when interventions are both easily extendable (Buildability) and require little effort to maintain user engagement (Continuity). Adherence to using an application builds on easy access to an intervention without encountering barriers (Accessibility) and providing reasons for users to stay engaged (Continuity). Instead of descriptively observing whether an app scales or is automated, we can concretely test: Is it accessible to those who need it? Can it be adapted without rewriting its architecture? Will people keep using it without repeated nudges?

### Limitations of the framework

Although designed to be broadly applicable, the ABC framework is not without limitations. One important caveat is that the model may not fit all digital health contexts equally. Interventions deeply tied to clinical environments, for instance, those integrated into electronic health records or delivered primarily through telehealth providers, often face constraints that limit their flexibility. In such settings, requirements for interoperability, compliance, and clinician oversight may reduce the extent to which interventions can be modular or user-driven. These constraints challenge the assumption that accessibility, buildability, and continuity can be optimized simultaneously and highlight that the ABC framework aligns more naturally with consumer-facing, app-based interventions than with tightly regulated clinical systems.

Another limitation lies in the “Buildable” dimension, which is deliberately broad, encompassing both developer and user chances of adapting the intervention. On the developer side, buildability can be defined concretely as referring to specific features like modular codebases, rapid iteration, and low-cost updates. On the user side, however, the concept remains less well defined. It also raises the open question of how much control users should even have in shaping their digital health experiences. While participatory design approaches stress the value of involving users in tailoring features to their contexts (Yardley et al., 2015), there is debate over whether users can always serve as accurate evaluators of their own needs and situations and how researchers and users might co-participate in implementing an intervention (Grüning et al., 2026).

This uncertainty is connected to findings from research on the description-experience gap, which shows that individuals often make different choices depending on whether information is learned from explicit descriptions or from lived experience. For example, people tend to overweight rare events when presented with descriptive information but underweight them when learning from experience, leading to systematic biases in decision-making (Garcia et al., 2021; Hertwig & Pleskac, 2018). This creates the challenge to help users balance their learnings from descriptive and experiential information to make them maximally accurate decision-makers when it comes to implementing their own interventions.

Lastly, the ABC framework deliberately emphasizes design-level factors as these have been rather underexplored and conceptualized in existing models. Through this focus, ABC pays less attention to systemic and contextual barriers such as digital literacy, health equity, and institutional adoption. Studies in implementation science show that these systemic barriers are decisive for the eventual success or failure of interventions as broader applications (Zamboni et al., 2019). These higher-level constraints should be considered along the ABC framework to obtain a holistic picture of what drives digital health innovation in the long term.

**Future research**

The ABC framework is intended as a practical design model and as such generates several promising avenues for future research. First, future work should investigate how the three dimensions of accessibility, buildability, and continuity manifest across different cultural, organizational, and technological contexts. Digital health interventions are increasingly deployed globally, yet user expectations, the affordances in health systems, and behavioral norms vary considerably. Understanding how ABC principles translate across these contexts may help identify both universal and context-specific design requirements to allow an application to be scalable, automated, and adhered to.

Second, the relationship between automation and user agency deserves closer examination. As digital interventions become increasingly capable of adapting and acting without direct user input, future research should explore how autonomy can be preserved while maintaining effectiveness. Concepts such as self-nudging (Herzog & Hertwig, 2025), which allow users to modify and shape intervention mechanisms themselves, may provide a fruitful direction for balancing the automation of digital health interventions with users remaining in control of their applications.

Third, hybrid intervention models that combine automated systems with human support are an important area for future development. That is, in some cases, the most scalable and effective interventions may be those that use automation to triage or augment human care, rather than replace it. Rather than viewing automation and human involvement as competing approaches, future research should investigate how they can be combined into one digital system. ABC can support this by framing digital components as part of a broader system, rather than a standalone solution of a fully independent application.

Lastly, the ABC framework itself requires empirical evaluation. Future studies should examine whether interventions designed according to ABC principles show higher scalability, automation and adherence, and, as a result, achieve greater long-term viability than interventions developed without such considerations. This creates opportunities for mixed-methods research, real-world implementation studies, longitudinal evaluations, and iterative development approaches embedded directly within intervention research. Such work may not just help corroborate the framework's design principles but also identify, which design principles contribute most strongly to successful real-world deployment, and what the framework has so far missed and could be refined by.

## Conclusion

This paper introduces the ABC framework, referring to Accessibility, Buildability, and Continuity, as a model for translating evidence digital health interventions into real-world applications, that is, interventions that are scalable, automated and adhered to by users. Where existing models outline high-level aspirations, ABC brings those concepts down to ground level, to the structure of apps, the architecture of features, and the interaction of these with user behavior. Additionally, we argue that the experienced need of users must drive the design of digital health interventions. The ABC framework helps shift focus from theoretical relevance to lived usefulness of digital health interventions. Ultimately, effective digital health is usable by many, adaptable by design, and engaging over time.

## Appendix

**Table A1**

*List of referenced smartphone applications and their descriptions.*

| Application | Description | Relevant functionalities referenced |
|---|---|---|
| **one sec** | Digital self-control and screen-time reduction app designed to interrupt habitual smartphone use. | App-specific interventions before opening selected apps; breathing pauses; customizable intervention types; per-app configuration; behavior analytics; time-saved feedback; dismissible interventions; background operation; passive behavioral support. |
| **Headspace** | Mindfulness and meditation application supporting stress reduction, wellbeing, and habit formation. | Free introductory content; rapid onboarding; guided meditations; immediate value delivery; streak tracking; progress feedback; long-term engagement mechanisms. |
| **Calm** | Meditation, sleep, and relaxation application focused on mental wellbeing and sleep improvement. | Free trial access; guided meditations; modular content library; sleep stories; sleep timer; passive value delivery during sleep; scalable content architecture. |
| **Strava** | Fitness tracking and social exercise platform for running, cycling, and related activities. | Activity tracking; progressive feature disclosure; user-generated route segments; leaderboards; social reinforcement ("kudos"); run clubs; privacy controls; API integrations; ecosystem connectivity with wearables and third-party services. |
| **MyFitnessPal** | Nutrition and health-tracking application focused on dietary monitoring and calorie management. | Large user-generated food database; crowd-sourced content creation; self-monitoring; scalable content generation through user contributions. |
| **Habitica** | Gamified habit and task management application that turns behavior change into role-playing game mechanics. | Social accountability systems; party/group challenges; peer reinforcement; collaborative goal achievement; optional social engagement. |